\documentstyle[aas2pp4]{article}

\slugcomment{}

\lefthead{}
\righthead{}

\begin{document}

\title{Simultaneous Multicolor Detection of Faint Galaxies in the Hubble Deep Field\footnote{
Based on observations made with the NASA/ESA
Hubble Space Telescope, obtained from the data archive at the
Space Telescope Science Institute, which is operated by the
Association of Universities for Research in Astronomy, Inc., under
cooperative agreement with the National Science Foundation}
}

\author{Alexander S. Szalay, Andrew J. Connolly and Gyula P. Szokoly}
\affil{Department of Physics and Astronomy,\\
The Johns Hopkins University, Baltimore MD 21218
}

\begin{abstract}
We present a novel way to detect objects when multiband images are
available. Typically, object detection is performed in one of the
available bands or on a somewhat arbitrarily co-added image. Our
technique provides an almost optimal way to use all the color
information available. We build up a composite image of the N
passbands where each pixel value corresponds to the probability that
the given pixel is just sky. By knowing the probability distribution
of sky pixels (a $\chi^2$ distribution with N degrees of freedom), the
data can be used to derive the distribution of pixels dominated by
object flux. From the two distributions an optimal segmentation
threshold can be determined. Clipping the probability image at this
threshold yields a mask, where pixels unlikely to be sky are
tagged. After using a standard connected-pixel criterion, the regions
of this mask define the detected objects.  Applying this technique to
the Hubble Deep Field data, we find that we can extend the detection
limit of the data below that possible using linearly co-added
images. We also discuss possible ways of enhancing object detection
probabilities for certain well defined classes of objects by using
various optimized linear combinations of the pixel fluxes (optimal
subspace filtering).

\end{abstract}


\section{Introduction}

Object detection, especially at faint flux limits, has always been
considered something of a ``black art''. This is despite the fact that
all techniques use the same basic steps, i.e.\ filtering of the images
with a PSF-like kernel, clipping of the filtered data using a
heuristically chosen threshold and then rejecting objects with too few
connected pixels (FOCAS --- Valdes 1982, Sextractor --- Bertin \&
Arnouts 1996). Each of these methods works well in the high
signal-to-noise regime, easily detecting all objects that lie
significantly above the sky noise. There is, however, a substantial
difference in the efficiency of the object detection process,
dependent on how we filter the images and what thresholds are applied
to identify objects as the detection threshold approaches the noise
level.

When multiband photometry is available, the problem of how to
systematically detect objects in all passbands becomes significantly
more complex. There are two basic approaches that are commonly used in
the astronomical literature. Object detection is undertaken in a
single passband (or a co-added image) with a matched aperture used to
measure the fluxes of the object in the additional passbands or object
detection is applied to each passband independently and the resulting
catalogs are merged to give a master list. Each of these approaches has
its own set of problems. Using a single image to derive a catalog can
miss objects with unusual colors or objects at the limit of detection
in a single passband (e.g.\ it is easy to imagine that if we set an
object detection limit at $5\sigma$ for each pixel in a single band we
will miss objects which would be a $4.5\sigma$ detection in each
individual passband but if we had considered all bands together we
would have accepted this pixel with a very high significance).
Detecting independently in each passband requires matching of objects
that may be detected in only one passband or be resolved into a
different number of components dependent on the passband the object was
observed in.

In this paper we attempt to deal with these detection problems in a
simple and objective fashion that segments images by taking into
account all the color information available. In Section 2 we describe
the statistical basis and the formal description of our
technique. Sections 3 contains the description of our technique as
applied to the Hubble Deep Field, the ultimate faint multicolor
dataset that has been a challenge for advanced processing
techniques. In Section 4 we discuss the results, and the use of
optimal subspace filtering to enhance the detection of fainter objects
of a specific color.

\section{Multicolor Segmentation of Object Pixel from Sky Noise}

\subsection{Transformation of the data into a probability image}

Consider a region on the sky that has been imaged in $N$ different
photometric bands all of which have been registered to a common scale
and size, thus every pixel contains $N$ numbers, the measured fluxes
$f_i, i=1..N$. We will further assume that the images are sky noise
limited, i.e. instrumental sources of pixel noise are irrelevant
compared to the photon noise of the sky background. In order to
proceed further we first determine the mean $\mu_i$ and the Gaussian
dispersion $\sigma_i$ of the sky in each band. This can be performed
using standard techniques, e.g.\ by studying the distribution of
pixels in the vicinity of the mode of the pixel values. In the next
step each image is transformed into one where the sky is a normal
Gaussian with zero mean and unit dispersion.  Each pixel is thus
described by the $N$-dimensional vector
\begin{equation} 
	g_i  =  { {f_i - \mu_i}\over \sigma_i}, \qquad i=1,...N.
\end{equation}
If there were no objects in the image, only sky, each component of
these vectors would be distributed as a normal Gaussian. The
probability distribution of $y$, the squared length of the vector,
\begin{equation}
	y = \sum_{i=1}^N g_i^2
\end{equation}
is then a $\chi^2$ with $N$ degrees of freedom and each value of $y$
is uniquely associated with the probability density,
\begin{equation}
	dP(y) = \frac{1}{2^{N/2} (N/2 - 1)!}\   e^{-y/2} y^{N/2 -1} dy.
\end{equation}

Since this is a monotonic function with respect to $y$, we can create
a new image consisting of the $\chi^2$ values. Any value of a pixel in
this image can be immediately interpreted as the cumulative
probability $P(>y)$ of how likely this pixel is drawn from the
sky distribution. This probability can be computed or read from a
standard table. In the actual ``probability''-image we will just use
the $y$ values for convenience since the mapping is monotonic.

If we have performed filtering on the original images (e.g.\ with a
PSF-like kernel), the pixel values may be correlated as $w_g(\theta)$,
even if the original sky pixels were not. It was shown (Adler 1981),
that in the special case of 2D random fields the $\chi^2$ field has a
correlation function $w_y(\theta) = 2N w_g(\theta)^2$, thus in the
filtered and non-filtered images we have an analytic understanding of
how the statistical properties behave under our null hypothesis (i.e.\
a purely sky noise image).

\subsection{Segmentation of the Probability Image, and the Optimal
Choice of Threshold} 

The next step is to consider the deviations from our null-hypothesis,
i.e. that all pixels are drawn from a Gaussian sky. This will be done
by clipping the $\chi^2$-image at a threshold high enough that a pixel
is unlikely to be drawn from the sky at a particular significance
level. Figure 1 shows the distribution of the $R=\sqrt{y}$ values
derived from the Hubble Deep Field (see Section 3). The long-tailed
contribution due to the presence of objects within the images is
clearly visible. In principle we could choose an arbitrary threshold
in the $y$ values to separate those pixels dominated by object flux
from the sky pixels. As we know the expected probability distribution
for the sky pixels (a $\chi^2$ distribution) we can approach this
problem in an objective fashion.

For small values of $y$ we know that the contribution to the flux in
each pixel will be due to the sky and that the distribution will be
that of a $\chi^2$. In Figure 1 the solid line shows the fitted
$\chi^2$ distribution and the dashed line the difference between the
$\chi^2$ and the observed distribution (i.e.\ the contribution due to
the objects).  The point where the two lines intersect determines the
so called ``optimal Bayes threshold'' (Fukunaga 1990), resulting in
the smallest possible total error, i.e.\ the sum of erroneously
tagging a sky pixel as an object, or missing an object pixel. If a
different 'cost' for the two kinds of errors is adopted, making one
more expensive, the optimal threshold can be computed from minimizing
that particular cost function. For example, we may want to set our
threshold much lower, at the expense of many false detections, in
order to find all the objects in the field.

Clipping at this threshold results in a binary mask image, where every
pixel is tagged with a value of 0, if it is consistent with sky, and 1
if it is unlikely to be sky. The next step in the processing is the
creation of contiguous regions. Objects which consist of too small a
number of pixels need to be rejected. In principle such a distribution
can be analytically calculated, at least in 2 dimensions, in the high
threshold approximation (Adler 1978), but we propose an empirical
approach: it is easy to generate entirely Gaussian random images, then
apply the same filtering, thresholding and segmentation as for the
data above. The region statistics computed from such images will then
yield the distribution for our null hypothesis. Subtracting this from
the data will give us the expected distribution of the object
sizes. An optimal threshold is then again determined from these two
distributions, as above.

\section{Faint Object Detection}

In the following section we describe the application of these
techniques to the multicolor $U_{300}$, $B_{450}$, $V_{606}$ and
$I_{814}$ photometry of the Hubble Deep Field (HDF; Williams et al.\
1996). To compare the $\chi^2$ image detection algorithm with the
standard methods we restrict our analysis to a single WFPC2 chip (chip
4) taken from Version 2 of the drizzled HDF data. We further restrict
the data we consider by excluding the outer 200 pixels from each image
to remove those regions of each frame where the noise is considerably
higher, due to the dithering pattern. The final frames comprise, for
each of the four passbands, 1781$\times$1741 images with 0.04 arcsec
pixels.

In the construction of these images they have already undergone a
fairly detailed reduction process, which includes compensation for bad
pixels and cosmic ray removal. They still require, however, additional
processing in order to derive a pixel set with more uniform statistical
properties. 

\subsection{Preprocessing of the images}
	
\subsubsection{Flattening the image}

If the mean sky background differs from zero across the frame (i.e.\
the sky has not been accurately subtracted from the images) the $y$
image will not approximate a $\chi^2$ distribution, even without the
presence of objects on a frame. In fact how well the $y$ image matches
a $\chi^2$ distribution, for low values of $y$, is a very sensitive
test of the background subtraction technique. We know from examining
the HDF data that they suffer from residual scattered light,
particularly in the $U_{300}$ passband (Williams et al.\ 1996). This
arises as an X pattern, due to the shadowing of the scattered light by
the mirror supports, superimposed on the background sky and occurs at
the few percent level. We must, therefore, fit and remove this
contribution before constructing the $y$ image.

As this is an additive contribution, it can be subtracted - its effect
on the level of sky noise is minimal. To create a smooth sky we derive
a mask, corresponding to a very generous cut of non-sky pixels. We set
our threshold to select the upper 16\% of pixels in a variance image
measured over a 3x3 pixel neighborhood. To safely stay away from the
wings of objects we grow this mask outwards by 5 pixels. These masked
pixels defined the object regions. Next, we grow these regions by an
additional 14 pixels. This provides an annulus of sky around each
object. Pixels within neighboring object regions that intersect these
annuli have been masked off. The mean value of the sky has been
computed within these annuli and the central regions corresponding to
objects filled-in with this value. Finally, we smooth this background
image image with a 46 pixel wide Gaussian filter and subtract it from
the original frame (creating images with zero mean sky).

\subsubsection{Normalizing to unit dispersion}

The final step in preprocessing the images is to convolve with a 3
pixel wide symmetric Gaussian kernel to perform the usual PSF
filtering. This step removes the arbitrariness of the location of
pixel boundaries, and increases the S/N of the data. The HDF images
have also a rather peculiar correlation due to the pattern of the
drizzling. By choosing a filter slightly broader than this width, we
are insensitive to the details of the drizzling.

We then build a histogram of the pixel values in each of the bands and
measure the dispersion of the sky. This is done by fitting an
inverted parabola to the log of the pixel histogram, using only the
data range around the modal value, extending slightly more to the
negative.  The fits were found to be excellent. The images were then
divided by the dispersion, resulting in zero mean, unit dispersion
Gaussian images.

\subsection{Building the $\chi^2$ image}

To build up the $\chi^2$ or $y$ image we sum the squares of the
individual pixel values in the four normalized images. From this we
construct the probability histogram (see Figure 1).  For the HDF data
we choose to consider the distribution of the square root, $R$ of the
$y$ values (i.e. $R = \sqrt{y}$). We do this because R corresponds to
a signal-to-noise statistic (as opposed to the squared signal-to-noise
of $y$) and can, therefore, be more naturally compared with the
standard techniques of thresholding images at a given multiple of the
sky noise. $R$ is the length of the $N$-dimensional pixel vector. The
normalized probability distribution for R, with 4 degrees of freedom,
corresponds to
\begin{equation}
	dP(R) = {1\over 2} R^3 e^{-R^2/2} dR.
\end{equation}

In Figure 1 we show the distribution of pixel values. As noted in
Section 2.2 the long positive tail is due to the contribution from
objects within the image frame. We fit the probability distribution
given in Equation 4 to the points below the peak of the histogram
(i.e. $R<2$) by varying the number of sky pixels as the only free
parameter. The fit is shown as a dotted line within Figure 1. The
difference between the fitted ``sky'' distribution and the observed
histogram is given by the dashed line. This represents the probability
distribution of pixels occupied by object flux. The intersection of
these two lines provides the Bayesian threshold (where the
incompleteness and contamination errors, at a pixel level, are
equal). We choose this crossover point ($R=3.73$) as the threshold
used in the image detection and analysis routines, in order to
demonstrate our ideas. This is a rather conservative cut, as we shall
see, since this corresponds to a 2.42$\sigma$ per pixel pixel
detection.

\subsection{Object detection}

Object detection is undertaken using Sextractor v2.0.8 (Bertin \&
Arnouts 1996). This version of Sextractor allows the detection and
analysis algorithms to be run on separate images (i.e.\ we can detect
objects on one frame and measure the photometric properties of these
detected objects from a separate image). We, therefore, use the $R$
image for detection and measure the fluxes and magnitudes for a fixed
set of apertures in the $U_{300}$, $B_{450}$, $V_{606}$ and $I_{814}$
passbands.  As all detections are undertaken using a single image this
procedure alleviates the task of matching catalogs from multiple
passbands (where images from different filters may be undetected in a
particular passband or broken up into multiple images).

The intersection of the sky and object distributions corresponds to an
$R$ value of $R=3.73$. This is equivalent to a probability of 0.9924
that a pixel value in the convolved image is not due to the
background sky. We adopt this relatively high probability because the
detected objects are unlikely to be contaminated by spurious sources
and, as the goal of this paper, is to demonstrate the applicability of
our new technique and to compare it with standard techniques, we do
not wish to push the initial application into the noise.

It should be noted that Sextractor, as with most object detection
packages, fits and subtracts a sky image before applying a detection
threshold. As the $\chi^2$ image is, by definition, a positive
definite quantity, subtracting a background will affect the detection
threshold. We have, therefore, modified the code to exclude background
subtraction for the detection image. To derive fluxes in the
photometric passbands for each detected object, matched aperture
photometry was then performed on the $U_{300}$, $B_{450}$, $V_{606}$
and $I_{814}$ images separately.

Of course, when searching for faint objects, one may abandon the
choice of the optimal Bayes classifier, rather one would set a much
lower detection threshold. Such a threshold will shift the balance
between the two errors, in essence we detect all the real objects, at
the expense of a lot of false detections. In such a case one can use
some more heuristic constraints to reduce the number of these spurious
sources. For example, one can require that in the detected objects all
of the fluxes be non-negative, or differ from zero by less than
$1\sigma$, or one can require that at least one of the bands has a
positive flux. Since the sky noise has a symmetry around zero in each
of the bands, one can subdivide our detections into $2^N$
`hyper-quadrants', according to the sign of their fluxes in each of
the $N$ bands. In the all-negative quadrant, all objects are false
detections.  Since there should be a similar number in the other
`hyper-quadrants', we can use the all-negatives as an estimator of the
false detections.  In practice, at the limit of the $R=3.73$
threshold, this filtering of the catalog makes little difference to
the number of detected sources.

\section{Discussion}

\subsection{Comparison of the $\chi^2$ detection to standard techniques}

As we have shown in Section 2 the $\chi^2$ detection algorithm
provides an almost optimal use of the multicolor information. We now
compare the results from the HDF data with the standard detection
techniques found in the literature. Specifically we consider three
cases, one where the initial detection is undertaken on the I band
image alone (using matched apertures on the other 3 passbands to
derive the multicolor photometry), second where we detect galaxies
using the sum of the V and I images (as was done for the HDF data in
Williams et al.\ (1996)) and a third case where we co-add all four
passbands weighted by the inverse of their sky noise (the weighting of
the passbands is given by the vector in Equation 5).

To provide a direct comparison between these differing techniques we
need to detect galaxies to a common threshold. The $R=3.73$ threshold
corresponds to a probability of 0.9924 that the flux in a pixel in the
convolved image exceeds that due to the sky alone (i.e.\ contains
object flux). For a Gaussian sky distribution with zero mean this is
equivalent to a 2.43$\sigma$ deviation.  We, therefore, convolve the I
and $V+I$ images with a kernel of 3 pixels FWHM (as was done for the
original images used to create the $\chi^2$) and use Sextractor to
derive an object catalog with a 2.43$\sigma$ threshold. Note that this
deliberately conservative threshold is substantially higher than that
applied to the HDF data by Williams et al.\ (1996).

Figure 2 shows the comparison between these different approaches (for
each of the four passbands), with the $\chi^2$ technique shown as a
long dashed line, the weighted image ($PC1$) as a short dashed line, the
$V+I$ detection as a solid line and the I detection as a long then
short-dashed line. For bright galaxies ($I<27.5$) each of the three
techniques find similar numbers of objects on a frame. As we move to
fainter magnitude limits the numbers of objects detected in the $PC1$,
$V+I$ and $\chi^2$ images clearly exceed those found in the I band
image. The magnitude limit for the detection of objects in the $PC1$ and
$V+I$ images are approximately 0.4 magnitudes fainter than that
determined from the I image. The comparable depths of the $PC1$ and $V+I$
catalogs is not surprising. The difference between the weight vector
for the $PC1$ image and that for the $V+I$ image is only 17
degrees. Therefore, the $V+I$ image provides essentially an optimal
linear combination of the four passbands (given the relative depths of
the data). It differs from the $PC1$ image only in that it is not
sensitive to very blue objects. The $\chi^2$ detection limit extends,
however, 0.4 magnitudes deeper than either of the $PC1$ and $V+I$ image.

The difference between the different detection techniques on the
histogram is striking. Since the thresholds were locked together
through the pixel tail probabilities, one obvious reason could be that
simply taking a lower threshold in the Gaussian detections would have
reproduced the result of the $\chi^2$ clipping. There will be
obviously some differences, due to the different behavior of the
excursion sets (regions consisting of pixels above a threshold) for
Gaussian and $\chi^2$ fields. Imagine an object in a single band image
consisting of $m$ pixels, all exactly at our selection threshold. The
sky noise will statistically move half the pixels below the detection
threshold, while the other half will remain above. For a $\chi^2$, in
the high threshold limit, approximately the same thing happens, but as
we approach lower and lower fluxes, it is more and more likely that
the effect of sky noise will be such, that the pixel remains above the
threshold, thus the destructive interference is much smaller, at the
expense of more false detections, of course.

Last, but not least it is also possible, that we simply detect objects
with such colors, that were undetected in the single (or co-added)
Gaussian band. This can be seen in Figure 3 where we show the
distribution of ($B-V$) colors of objects found in the $PC1$ and
$\chi^2$ images. The solid histogram shows the color distribution for
objects selected from the $PC1$ images and the open histogram the color
distribution from the $\chi^2$ image. The weighting of the $PC1$
selection preferentially selects objects with intrinsically redder
colors (with a mean color of $B-V = 0.46$). The $\chi^2$ image, by
contrast, has a more symmetric color distribution with a mean color
$B-V = 0.37$. Also note, that the scatter in the color distribution is
considerably broader for the $\chi^2$ detections.

We show that this is a significant effect, by splitting the objects
into two categories: those detected in both the $PC1$ and $R$ images and
those detected in $R$ only (there were no objects found in $PC1$ but
not in $R$).  Next, we create a histogram of the number of these
detections versus the peak of the $R$ value inside the object. In
Figure 4 we show this distribution with those objects found in the
$PC1$ and $R$ images shown by a solid line and those found in $R$
alone shown by a dashed line. We can see, that there are objects,
un-detected in $PC1$, that have a very significant value of $R$.

In Figure 5 we show the $B$, $V$ and $I$ images of 4 of those objects
which had significant $R$ values ($R>10$) but were not detected in the
$PC1$ image. Together with the optical passbands we show the $R$ image
for each object. It is clear from these images that those objects
selected in the $R$ frame but not in $PC1$ are predominantly bright in a
single passband (possibly due to strong emission lines). A linear
combination of the available passbands (whether the $V+I$ combination
or the $PC1$ weighted sum) does not maximize the signal-to-noise of the
resultant image for these galaxies with unusual colors. The $R$ image
does, however, provide an optimal combination of the available
passbands and can be used to derive a sample of galaxies that is not
biased towards a specific type while maximizing the information
present in the multicolor data.


\subsection{Optimal Detection of Objects of Specific Color}

The $\chi^2$ technique we have described above is the optimal way for
detecting objects within an image with no prior information. In this
section we outline a possible extension to our technique given a model
for the colors of the objects we expect to detect. 

In our four-dimensional flux system (for the HDF) every pixel is a 4D
vector. Sky pixels are uniformly scattered, all directions are equally
probable. Objects or pixels of a specific color on the other hand all
point to the same direction, denoted by $\vec n$. Since the noise with
our scaling is spherically symmetric, the component of the noise in
this direction is also a normal Gaussian. Thus by ignoring all the
other components, i.e. taking a linear combination of the different
images along the $\vec n$ direction, we discard 3 of the 4 noise
components, an effective increase in our signal-to-noise of 2. Thus we
are keeping as much of the signal as possible, while including as
little of the noise as possible. Similarly, we can define a 2
dimensional subset, where most of the objects lie, and we can create a
$\chi^2$ image in that 2D subset of the 4D color space. These
techniques are called, for the obvious reason, ``Optimal Subspace
Filtering''. In order to determine the optimal directions, we have
performed a Principal Component Analysis for all the pixels which were
above our threshold of $3.73\sigma$. The resulting vector is the
following:
\begin{equation}
	{\vec n} = (0.075, 0.29, 0.66, 0.69)
\end{equation}
Note, that this is quite near to the $V+I$, the cosine of the angle
is 0.95. This would be the preferred direction to use to go after
objects of `typical' colors, but it would have the danger of losing
faint objects at the extreme ends of the color distribution.

If we know colors of our preferred objects a priori, i.e. we are
focusing on a narrow range of galaxy colors, we can use the linear
combination along the corresponding direction to enhance our
detection. In such a way one can develop particularly sensitive
detections of specific classes of object.

\subsection{Conclusions}

We have formulated an object detection algorithm that is capable of
using all the multicolor information available in a given field.
Applying this algorithm to the Hubble Deep Field we have shown that it
performs substantially better than standard image analysis techniques.
Given a certain probability threshold, we find that the $\chi^2$-based
detection technique can identify faint objects of unusual colors,
where detections on single co-added images may fail. We outline several
heuristics on how false detections can be reduced, should one adopt a
very low detection threshold, using the isotropy properties of the
multidimensional sky-noise and we discuss how to optimize these
detections for objects of specific colors. Given the increasing
commonality of large multicolor photometric surveys we believe that
the adoption of techniques such as we describe here would
significantly improve the resulting object catalogs.

\acknowledgements 
We would like to thank the HDF team for making their data so rapidly
available to the public. Most of our image processing has been
implemented using KBVision, helpful advice from Pete Eggleston (AAI)
is happily acknowledged in this task.

\begin{figure}
\caption{ The distribution of pixels in the $R$-image is shown for
chip 4 of the HDF data set. The solid line shows expected distribution
for a $\chi^2$ with 4 degrees of freedom. The longer positive tail is
due to the contribution of pixels containing object flux. When we
subtract the expected distribution for the sky we derive the dashed
line (showing the distribution of object pixels). Where these two
distributions cross is the Bayes threshold where the error due to
misclassifying object pixels as sky and sky pixels as objects is
equal. For chip 4 this threshold is $R=3.73$ corresponding to a
probability of 0.9924 that a pixel is not drawn from the sky
distribution.}
\end{figure}

\begin{figure}
\caption{The number counts derived from Chip 4 of the HDF data sets
for the $U_{300}$, $B_{450}$, $V_{606}$ and $I_{814}$ passbands. The
long-dashed line shows the number counts derived using the $\chi^2$
image for detection, the short-dashed line the number counts from the
weighted sum of all passbands ($PC1$), the solid line the counts from
the $V+I$ images and the long-short dashed line from the I image. Each
detection technique has been applied at the same probability limit
(0.9924 that the pixel is not drawn from the sky).  In the bright
limit $I<27.5$ all detection techniques find similar numbers of
objects. At fainter magnitude limits the $PC1$ and $V+I$ detection
exceed that of the I image by approximately 0.4 magnitudes and the
$\chi^2$ image exceeds these by a further 0.4 magnitudes.}
\end{figure}

\begin{figure}
\caption{The color distribution $B_{450} - V_{606}$ of objects
detected in the HDF images. The filled histogram is derived from the
objects detected from the $PC1$ image and the unfilled histogram from
the $\chi^2$ image. The $PC1$ image detection results in a color
distribution that is skewed to the red in $B_{450} - V_{606}$
color. The $\chi^2$ image provides a more Gaussian distribution of
objects centered on a mean color of $B_{450} - V_{606} = 0.37$.}
\end{figure}

\begin{figure}
\caption{The solid line indicates the distribution of objects detected
in the $PC1$ image, as a function of the peak $R$ value within the
object, while the dashed line corresponds to objects detected in the
$\chi^2$ image, {\it and} undetected in the $PC1$. Note, that there
are several objects at rather high $R$ values.
}
\end{figure}

\begin{figure}
\caption{Cutouts of a few objects with odd colors, which were detected
in the $\chi^2$ image, but remained undetected in $PC1$. The leftmost
bitmap shows the $R$ image (the distance from the origin in multicolor
S/N space), while the next three show the flattened $B, V$ and $I$
images. Each detected object appears to be bright in a single passband
(possibly due to strong emission lines) and weak in the other two. All
of these bands contribute to the signal in the $R$ image. Note, how
obvious the objects look in the $R$ image, compared to the other
three. The 4 objects shown can be identified in the original drizzled
WFPC2 chip 4 images at coordinates (a) $x=1958$ $y=1016$, (b) $x=342$
$y=1649$ (c) $x=1891$ $y=1062$ (d) $x=1413$ $y=1586 $. }
\end{figure}

\clearpage

\end{document}